\documentclass[prb,amsmath,twocolumn,amssymb]{revtex4-1}
\usepackage{times}
\usepackage{graphicx}
\usepackage{amsmath}
\usepackage{amssymb}
\usepackage{natbib}
\usepackage{bm}

\newcommand{\specialcell}[2][c]{%
  \begin{tabular}[#1]{@{}c@{}}#2\end{tabular}}

\def\lsim{\lower -0.3ex \hbox{$<$} \kern -0.75em \lower 0.7ex \hbox{$\sim$}}
\def\gsim{\lower -0.3ex \hbox{$>$} \kern -0.75em \lower 0.7ex \hbox{$\sim$}}

\begin{document}
\title{Tunable symmetry breaking and helical edge transport in a graphene quantum spin Hall state}

\author{A. F. Young$^{1,\dagger,\ast}$, J. D. Sanchez-Yamagishi$^{1,\dagger,\ast}$, B. Hunt$^{1,\dagger,\ast}$,S. H. Choi$^1$,K. Watanabe$^2$,T. Taniguchi$^2$,R. C. Ashoori$^1$ \& P. Jarillo-Herrero$^1$\\
\normalsize{$^{1}$Department of Physics, Massachusetts Institute of Technology}
\normalsize{Cambridge, MA, USA}\\
\normalsize{$^{2}$Advanced Materials Laboratory, National Institute for
Materials Science,}
\normalsize{Tsukuba, Japan}\\
\normalsize{$^\dagger$These authors contributed equally to the work.}
\normalsize{$^\ast$afy@mit.edu,benhunt@mit.edu,jdsy@mit.edu}
}
\begin{abstract}
Low-dimensional electronic systems have traditionally been obtained by electrostatically confining electrons, either in heterostructures or in intrinsically nanoscale materials such as single molecules, nanowires, and graphene.
Recently, a new paradigm has emerged with the advent of symmetry-protected surface states on the boundary of topological insulators, enabling the creation of electronic systems with novel properties. For example, time reversal symmetry (TRS) endows the massless charge carriers on the surface of a three-dimensional topological insulator with helicity, locking the orientation of their spin relative to their momentum\cite{Hasan2010,Qi2011}.  Weakly breaking this symmetry generates a gap on the surface,\cite{Chen2010} resulting in charge carriers with finite effective mass and exotic spin textures\cite{Xu2012}.  Analogous manipulations of the one-dimensional boundary states of a two-dimensional topological insulator are also possible, but have yet to be observed in the leading candidate materials\cite{Konig2007,Du2013}.  Here, we demonstrate experimentally that charge neutral monolayer graphene displays a new type of quantum spin Hall (QSH) effect\cite{Abanin2006,Fertig2006}, previously thought to exist only in time reversal invariant topological insulators\cite{Kane2005,Bernevig2006,Bernevig2006a,Konig2007}, when it is subjected to a very large magnetic field angled with respect to the graphene plane.   Unlike in the TRS case\cite{Kane2005,Bernevig2006,Konig2007}, the QSH presented here is protected by a spin-rotation symmetry that emerges as electron spins in a half-filled Landau level are polarized by the large in-plane magnetic field.  The properties of the resulting helical edge states can be modulated by balancing the applied field against an intrinsic antiferromagnetic instability\cite{Herbut2007,Jung2009,Kharitonov2011}, which tends to spontaneously break the spin-rotation symmetry.  In the resulting canted antiferromagnetic (CAF) state, we observe transport signatures of gapped edge states, which constitute a new kind of one-dimensional electronic system with tunable band gap and associated spin-texture\cite{Kharitonov2012}.
\end{abstract}
\maketitle

In the integer quantum Hall effect, the topology of the bulk Landau level (LL) energy bands\cite{Thouless1982} requires the existence of gapless edge states at any interface with the vacuum.
The metrological precision of the Hall quantization can be traced to the inability of these edge states to backscatter due to the physical separation of modes with opposite momentum by the insulating sample bulk\cite{Halperin1982}.
In contrast, counterpropagating boundary states in a symmetry-protected topological (SPT) insulator coexist spatially but are prevented from backscattering by a symmetry of the experimental system\cite{Qi2011,Hasan2010}.
The local symmetry that protects transport in SPT surface states is unlikely to be as robust as the inherently nonlocal physical separation that protects the quantum Hall effect. However, it enables the creation of new electronic systems in which momentum and some quantum number such as spin are coupled, potentially leading to devices with new functionality.
Most experimentally realized SPT phases are based on TRS, with counterpropagating states protected from intermixing by the Kramers degeneracy.  However, intensive efforts are underway to search for topological phases protected by other symmetries that can be realized or engineered in new experimental systems.

	\begin{figure*}[ht!]
	\begin{center}
 \includegraphics[width=6.5 in]{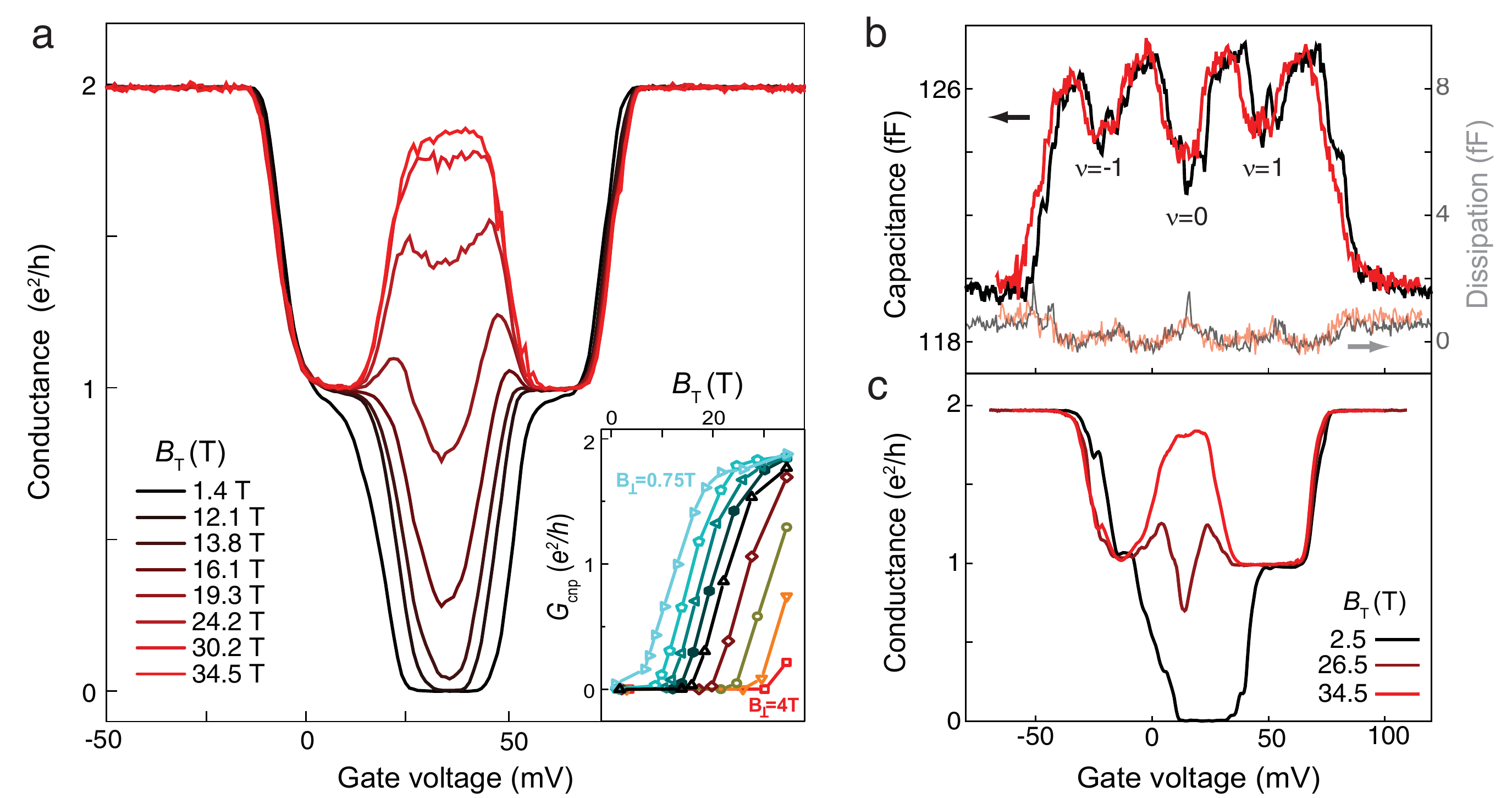}
		\caption{\textbf{Quantum spin Hall state in monolayer graphene in extreme tilted magnetic fields.} \textbf{a}, Conductance of device A at $B_\perp$=1.4T for different values of $B_T$.  As $B_T$ increases, the insulating state at $\nu$=0 is gradually replaced by a high conductance state, with an accompanying inversion of the sign of $\partial G_{cnp}/\partial T$ (Extended Data Figure 1). Inset: $G_{cnp}$ as a function of $B_T$ for Device A.  Left to right: $B_\perp$=0.75 (cyan), 1.0, 1.4, 1.6, 2.0, 2.5, 3.0, and 4.0~T.
\textbf{b,} Capacitance (dark lines) and  dissipation (faded lines) of device B at $B_\perp$=2.5T. The low dissipation confirms that the measurements are in the low-frequency limit, so that the dips in capacitance can be safely interpreted as corresponding to incompressible states.  \textbf{c,} Conductance under the same conditions.  The absence of a detectable change in capacitance, even as the two-terminal conductance undergoes a transition from an insulating to a metallic state suggests that the conductance transition is due to the emergence of gapless edge states.}
		\label{fig1}
	\end{center}
\end{figure*}

Our approach is inspired by the similarity between the TRS QSH effect and overlapping electron- and hole-like copies of the quantum Hall effect, with the two copies having opposite spin polarization.  This state is protected by spin conservation rather than the orthogonality of states in a Kramers doublet\cite{Kane2005}, as with the TRS QSH observed in strong spin-orbit systems.  Nevertheless, it is expected to reproduce the characteristic experimental signatures of the TRS QSH, with gapless helical edge states enclosing an insulating bulk\cite{Abanin2006,Fertig2006}. Two requirements are necessary for realizing such a QSH state.  First, the spin-orbit coupling must be weak, so that spin remains a good quantum number. Second, the energy gap between electron- and hole-like Landau levels must be small enough to be invertible by the Zeeman splitting.  Both of these conditions are met in graphene, a gapless semimetal with very weak spin-orbit coupling\cite{Min2006}.  The graphene LL structure is characterized by the existence of a fourfold spin- and valley-degenerate LL at zero energy(zLL)\cite{Semenoff1984}.  Near the sample boundary, the zLL splits into one positively dispersing (electron-like) and one negatively dispersing (hole-like) mode per spin projection.  Consequently, a spin-symmetry protected QSH effect is expected when the spin degeneracy is lifted by an external magnetic field, resulting in a bulk energy gap at charge neutrality and electron-like and hole-like states with opposite spin polarization that cross at the sample edge\cite{Abanin2006,Fertig2006}.

\begin{figure}[ht!]
\begin{center}
\includegraphics[width=3 in]{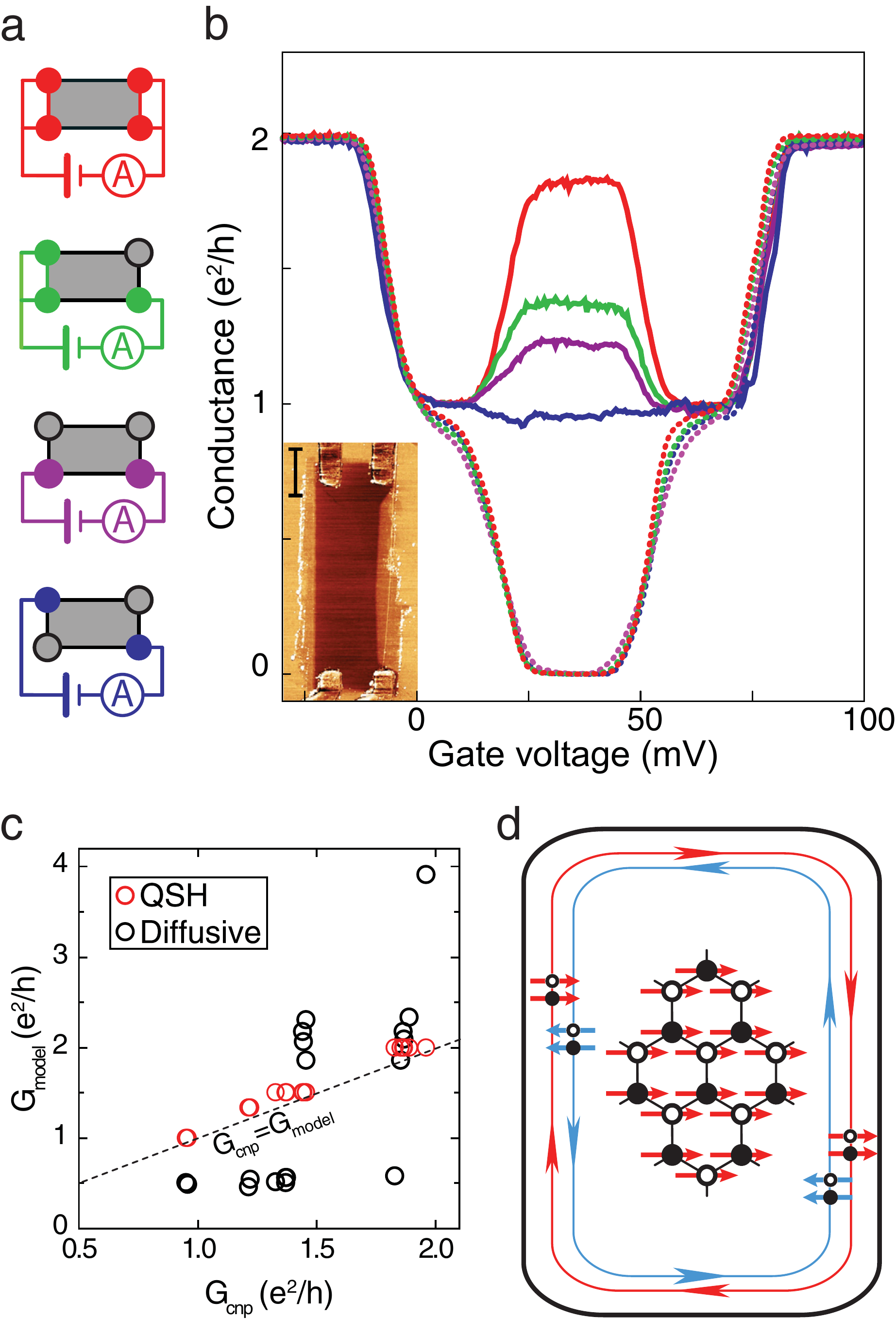}
\caption{\textbf{Nonlocal two-terminal transport in the quantum spin Hall regime.}  \textbf{a,}  Schematic of four distinct two-terminal measurement topologies available in a four-terminal device. Hollow circles indicate floating contacts while filled, colored circles indicate measurement contacts.  Each variation probes two parallel conductance paths between the measurement contacts with a variable number of segments on each path, indicated by black edges.
\textbf{b}, Two-terminal conductance measurements of Device A for $B_\perp$=1.4T, color-coded to the four different measurement configurations.  Dashed curves are taken at $B_T$=1.4T; solid curves are taken at $B_T$=34.5T (QSH regime).  In the QSH regime, $G_{cnp}$ depends strongly on the number of floating contacts. Inset: AFM phase micrograph of Device A. Scale bar: 1 micron.
\textbf{c}, $G_{cnp}$ for eighteen different contact configurations based on cyclic permutations of the topologies shown in \textbf{a}.  Data are plotted against two model fits.  In a numerical simulation based on a diffusive model (black circles), the graphene flake was assumed to be a bulk conductor with the conductivity left as a fitting parameter ($\sigma=3.25 e^2/h$ for the best fit).  The QSH model is Eq. 1, and has no fitting parameters. The dashed line indicates a perfect fit of data to model.  Note that the measured $G_{cnp}$ never reaches the value predicted by the QSH model, indicating either contact resistance or finite backscattering between the helical edge states.
\textbf{d}, Schematic of bulk order and edge state spin texture in the fully polarized QSH regime.  Arrows indicate the projection of the electron spin on a particular sublattice, with the two sublattices indicated by hollow and filled circles.  The edge state wavefunctions are evenly distributed on the two sublattices and have opposite spin polarization, at least for an idealized armchair edge\cite{Kharitonov2012}.
}
				\label{fig-nonlocal}
	\end{center}
\end{figure}

\begin{figure*}[ht!]
	\begin{center}
 \includegraphics[width=6.5 in]{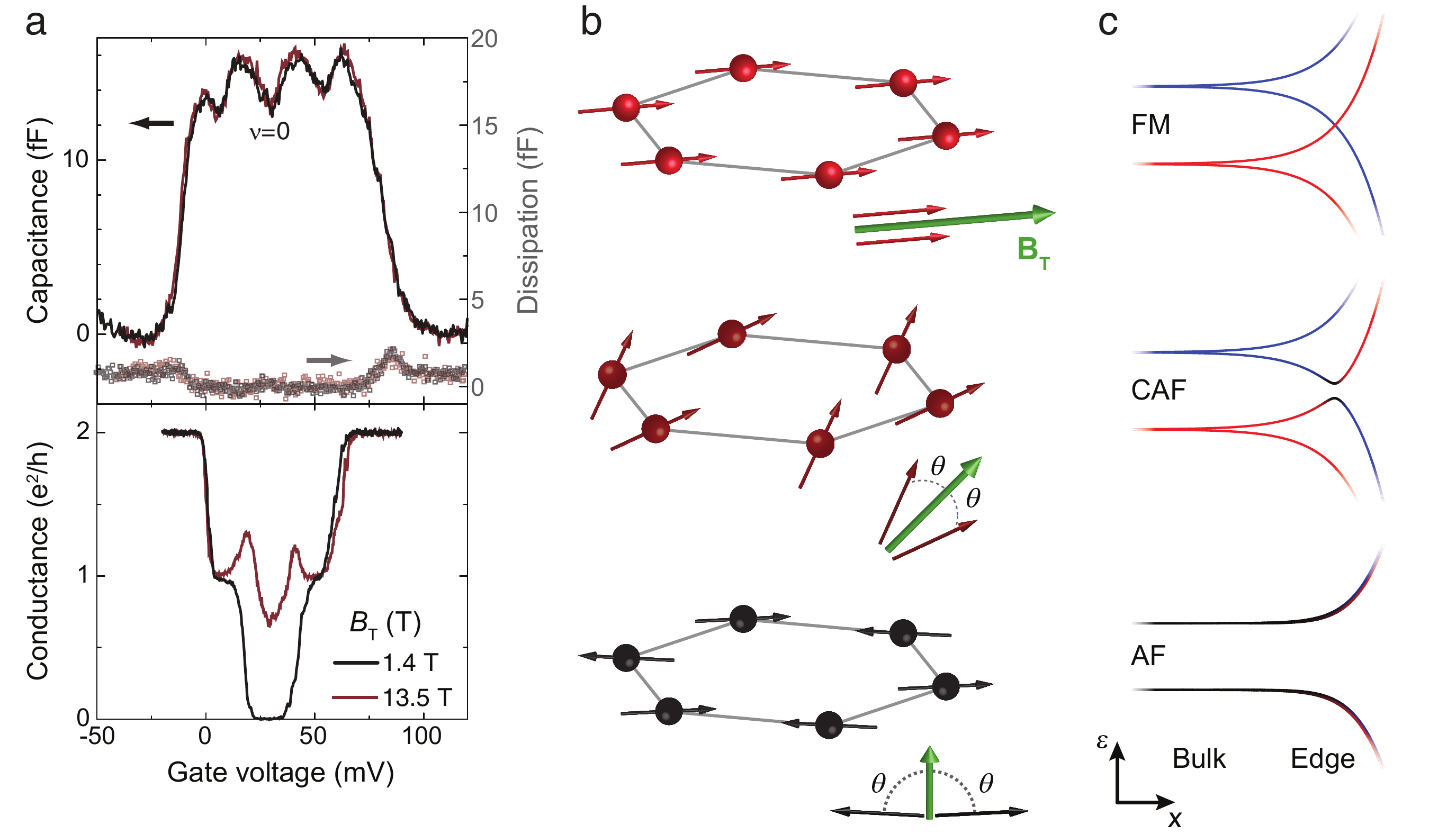}
\caption{\textbf{Symmetry-driven quantum phase transition.} \textbf{a} Capacitance (top) and conductance (bottom) of device A at $B_\perp$=1.1T.
The central dip in capacitance does not change with $B_T$ at any point during the transition, implying that the bulk gap does not close.
\textbf{b}, Bulk spin order in the three transition regimes. The balls and arrows are schematic representations of the spin and sublattice texture of the ground state wavefunctions and do not represent individual electrons; the electron density within the ZLL at $\nu$=0 is two electrons per cyclotron guiding center. The insets in \textbf{b} show details of the relative alignment of the electron spins on the two sublattices.
At large $B_T$, the bulk electron spins are aligned with the field (top panel), resulting in an emergent U(1) spin-rotation symmetry in the plane perpendicular to $B_T$.   As the total magnetic field is reduced below some critical value (with $B_\perp$ held constant), the spins on opposite sublattices cant with respect to each other while maintaining a net polarization in the direction of $B_T$ (middle panel).  This state spontaneously breaks the U(1) symmetry, rendering local rotations of the electron spins energetically costly. At pure perpendicular fields (bottom panel), the valley isospin anisotropy energy overwhelms the Zeeman energy and the canting angle $\theta$ is close to $90^\circ$, defining a state with antiferromagnetic order.
\textbf{c}, Low energy band structure in the three phases\cite{Kharitonov2012}. $\epsilon$ is the energy and $x$ is the in-plane coordinate perpendicular to the physical edge of the sample.  The intermediate CAF phase smoothly interpolates between the gapless edge states of the QSH phase (top panel) and the gapped edge of the perpendicular field phase (bottom panel) without closing the bulk gap.  Colors indicate spin texture of the bands projected onto the magnetic field direction, with red corresponding to aligned, blue antialigned, and black zero net spin along the field direction.  
}
		\label{fig-phasetransition}
	\end{center}
\end{figure*}

Experimentally, charge neutral monolayer graphene does not exhibit the expected phenomenology of the QSH effect, becoming strongly insulating instead at high magnetic fields\cite{Checkelsky2008}.
While the precise nature of this insulating state has remained elusive, its origin can be traced to the strong Coulomb interactions within the graphene zLL.
At integer filling factors, $\nu$, the Coulomb energy is minimized by forming antisymmetric orbital wavefunctions, forcing the combined spin/valley isospin part of the wavefunction to be symmetric. The resulting possible ground states lie on a degenerate manifold of states fully polarized in the approximately SU(4)-symmetric isospin space\cite{Yang2006}, encompassing a variety of different spin- and valley- orders.  In the real experimental system, the state at any given filling factor (such as $\nu=0$) is determined by the competition between SU(4) symmetry-breaking effects.  The most obvious such anisotropy is the Zeeman effect, which naturally favors a spin-polarized state, but the sublattice structure of the zLL adds additional interaction anisotropies\cite{Alicea2006} that can favor spin unpolarized ground states characterized by lattice scale spin- or charge-density wave order\cite{Herbut2007,Jung2009,Nomura2009,Kharitonov2011}.  This interplay can be probed experimentally by changing the in-plane component of magnetic field, which changes the Zeeman energy but does not affect orbital energies, and previous observations indeed confirm that the state responsible for the $\nu$=0 insulator is spin unpolarized\cite{Young2012}.  However, the spin polarized QSH state can be expected to emerge for sufficiently large in-plane field, manifesting as an incompressible conducting state at charge neutrality.

Figure 1a shows two terminal conductance measurements of a high quality graphene device fabricated on a thin hBN substrate, which itself sits atop a graphite local gate. As the total magnetic field ($B_T$) is increased with $B_\perp$ held constant, the initially low charge neutrality point conductance ($G_{cnp}$) increases steadily before finally saturating at $G~\sim~$1.8 $e^2/h$ for the largest total field applied.  Evidence for a similar transition was recently reported in bilayer graphene\cite{Maher2013}. To distinguish the role of the edges and the bulk in this conductance transition, we also measure the capacitance between the graphene and the graphite back gate under similar conditions.  Capacitance ($C$) measurements serve as a probe of the bulk density of states ($D$) via $C^{-1}=C_G^{-1}+(A e^2D)^{-1}$, where $C_G$ is the geometric capacitance and $A$ is the sample area. Simultaneous capacitance and transport measurements from a second graphene device show that quantized Hall states within the zLL at $\nu=0$ and $\nu=\pm1$ are associated with minima in the density of states (Fig. \ref{fig1}b-c).  As the total field is increased, the capacitance dip at $\nu=0$ remains unaltered even as conductance increases by several orders of magnitude. This implies that the high field $\nu=0$ state has an incompressible bulk, consistent with the hypothesis of a ferromagnetic QSH state with conducting edge states and a bulk gap.

We probe the nature of the edge states through nonlocal transport measurements in which floating contacts are added along the sample edges\cite{Roth2009}.
Unlike the chiral edge of a quantum Hall state, which carries current in only one direction, the QSH edge can carry current either way, with backscattering suppressed by the conservation of spin within the helical edge sates.  Because the carriers do not maintain their spin coherence within a metal contact, contacts equilibrate the counterpropagating states so that each length of QSH edge between contacts must be considered as a single $h/e^2$ resistor.
The two-terminal conductance results from the parallel addition of the two edges connecting the measurement probes,
\begin{equation}
G=\frac{e^2}{h} \left( \frac{1}{N_1+1}+\frac{1}{N_2+1} \right),
\label{GNL}
\end{equation}
where $N_1$ and $N_2$ are the number of floating contacts along each edge.
Figure \ref{fig-nonlocal}b shows the results of nonlocal two-terminal conductance measurements for the four distinct two-terminal measurement geometries available in a four-terminal device (Fig. 2a).  Repeating the measurement for 18 cyclic permutations of the available contact configurations, we find that the results are well fit by the simple model of Eq. \ref{GNL} (Fig. \ref{fig-nonlocal}c) despite large variations in the effective bulk aspect ratio.  Notably, $G_{cnp}$ is always less than the value expected from the QSH model, suggesting some small but finite amount of backscattering or contact resistance.  The combination of bulk incompressibility and nonlocal transport signatures of counterpropagating edge states lead us to conclude that the high field metallic state observed indeed displays a QSH effect.

\begin{figure*}[ht!]
\begin{center}
 \includegraphics[width=6.5 in]{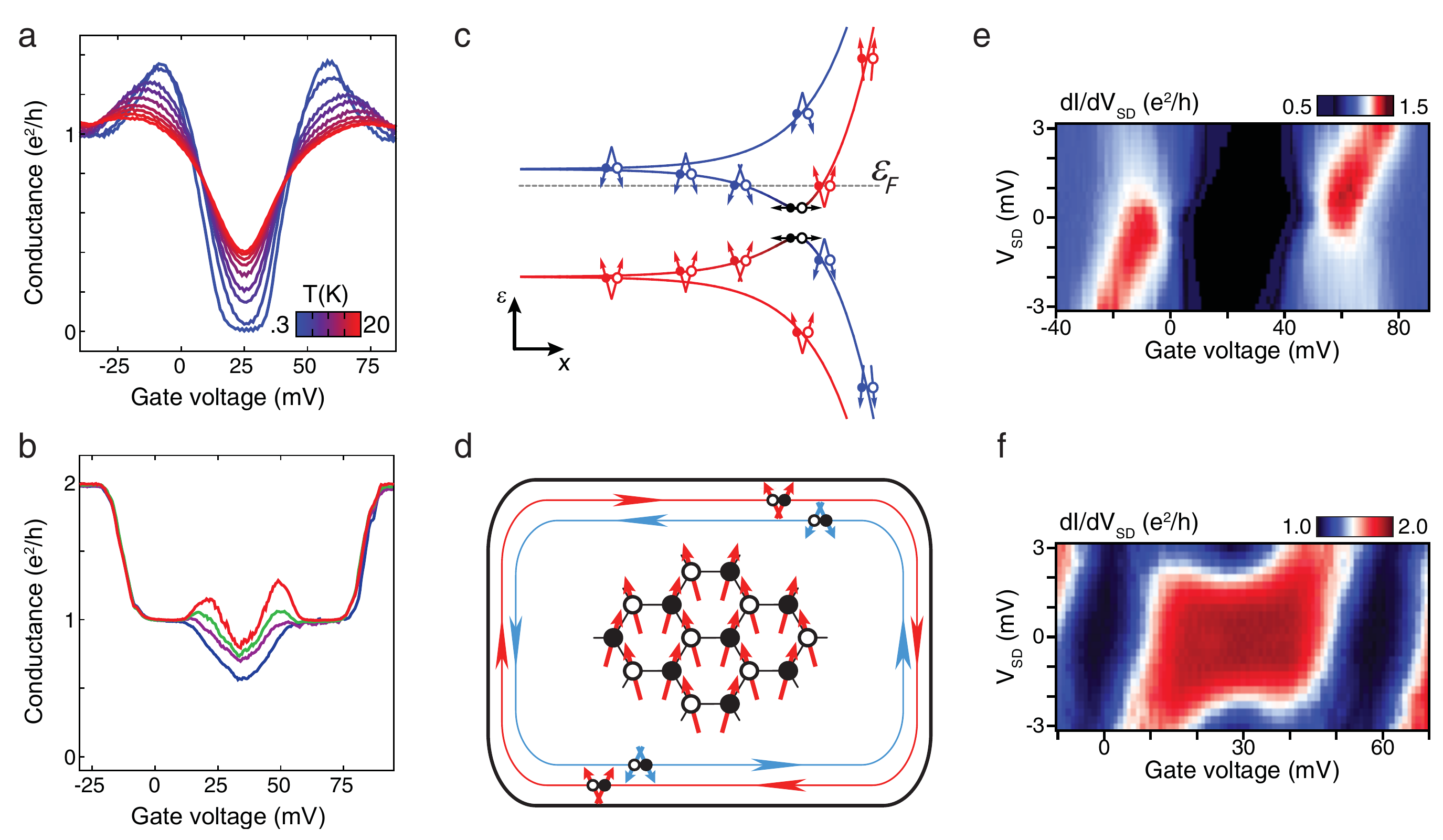}
\caption{
\textbf{Spin textured edge states of the CAF phase.}
\textbf{a,} Temperature dependence in the intermediate field regime for device C at $B_\perp$=5.9T and $B_T$=45.0T.  The conductance peaks shows a metallic temperature coefficient, while the state at charge neutrality remains insulating.
\textbf{b,} Nonlocal two-terminal conductance of device A at $B_\perp$=1.6T and $B_T$=26.1T. Color coding indicates contact geometry following the scheme in Figure \ref{fig-nonlocal}a.  The height of the conductance peaks depends strongly on the configuration of floating contacts, indicating their origin in the gapped, counterpropagating edge states of the CAF phase.
\textbf{c,} Schematic band diagram, including spin order, of the CAF edge states.  For the electron and hole bands nearest to zero energy, the canting angle inverts near the sample edge, leading to counterpropagating edge states with inverted CAF spin texture.  The dashed gray line indicates the Fermi energy, $\epsilon_F$, in the regime corresponding to one of the conductance peaks.
\textbf{d},  Schematic of bulk order and edge state spin texture in the CAF regime, following the convention of Fig. 2d.
\textbf{e}, Differential conductance, $dI/dV_{SD}$, of device C in the QSH regime ($B_\perp$=2.7~T, $B_T$=45.0~T) in units of $e^2/h$. A constant source-drain voltage, $V_{SD}$, along with a 100$\mu$V, 313 Hz excitation voltage, are applied to one contact and the AC current measured through the second, grounded contact.
\textbf{f}, $dI/dV_{SD}$ of device C in the CAF regime ($B_\perp$=5.9T, $B_T$=45.0T) in units of $e^2/h$.  In both \textbf{e} and \textbf{f},  a symmetry is observed upon reversing both $V_{SD}$ and carrier polarity.
}		
		\label{fig-CAFedgestates}
	\end{center}
\end{figure*}

The QSH state realized here is equivalent to two copies of the quantum Hall effect, protected from mixing by the U(1) symmetry of spin rotations in the plane perpendicular to the magnetic field.  As such, it constitutes a topologically nontrivial state that is clearly distinct in its edge state properties from the insulating state at fully perpendicular field.  What kind of transition connects the two states?  Capacitance measurements in the intermediate conductance regime reveal that the bulk gap does not close as the total field is increased (Fig. \ref{fig-phasetransition}a).  This rules out a conventional topological phase transition, in which case the bulk gap is required to close\cite{Buttner2011}; the transition must thus occur by breaking the spin-symmetry on which the QSH effect relies.  In fact, a canted antiferromagnetic state (Fig. \ref{fig-phasetransition}b) that spontaneously breaks this symmetry is among the theoretically allowed $\nu=0$ states\cite{Herbut2007,Jung2009,Kharitonov2011}. Within this scenario, the canting angle is controlled by the ratio of the Zeeman energy, $g\mu_B B_T$, and the antiferromagnetic exchange coupling, which depends only on $B_\perp$.  The observed conductance transition results from the edge gap closing (Fig. \ref{fig-phasetransition}c) as the spins on the two graphene sublattices are slowly canted by the in-plane magnetic field, with the fully polarized QSH state emerging above a critical value of $B_T$\cite{Kharitonov2012}. In the language of SPT insulators, the antiferromagnetic instability breaks the spin symmetry below this critical field, allowing the counterpropagating edge states to backscatter and acquire a gap.

Experimentally, the sub-critical field regime is characterized by high conductance peaks appearing symmetrically between $\nu=0$ and $\nu=\pm1$.  We observe $G>e^2/h$ peaks in many samples with widely varying aspect ratios, which is inconsistent with diffusive bulk transport in a compressible Landau level\cite{Abanin2008}.  Variable temperature measurements indicate that the peaks are metallic, even when the state at $\nu=0$ is still strongly insulating (Fig. 4a).  Moreover, the peaks exhibit the nonlocal transport behavior of counterpropagating edge states (Fig. 4b); in particular the peak conductance is always strictly less than $e^2/h$ when the two edges are interrupted by a floating contact.  These results indicate that the conductance peaks are due to edge state transport in the canted antiferromagnetic state.  The high conductance of these edge states, despite proximity to the strongly disordered etched graphene edge, implies that backscattering is at least partially suppressed.  This is consistent with the the theory of the CAF band structure shown in Fig. 4c-d\cite{Kharitonov2012}, in which the interpolation between the gapless QSH and gapped AF edge structure is achieved via a new kind of one-dimensional edge state in which counterpropagating modes have oppositely-canted AF spin texture.  Interestingly, existing theories of the CAF state are only rigorously applicable to the zero carrier density regime, in which case the CAF edge modes exist as excited states.  The fact that we can access the CAF edge states via gating is somewhat surprising, as it implies that this spectrum is stable to small populations of the edge bands.

Questions remain about the precise nature of the QSH and CAF boundary modes.  Experimentally, $G_{cnp}$ never reaches 2$e^2/h$ even at the highest values of $B_T$, despite some of the devices showing a flat plateau around charge neutrality.  Naively, backscattering within the QSH edge mode requires flipping an electron spin, for example by magnetic impurities, although such a process should be energetically unfavorable at high magnetic fields.  More trivially, we cannot exclude that weakly conducting charge puddles connect the two edges (but not source and drain contacts), leading to backscattering across the bulk in the QSH regime.  Spin-orbit effects may also play a role by spoiling the spin-symmetry upon which the helical edge states rely.  While the intrinsic spin-orbit coupling in graphene is thought to be weak\cite{Min2006}, the helical states may be uniquely sensitive to spin relaxation.  Alternatively, the large Rashba-type spin-orbit coupling induced in the graphene under the gold contacts\cite{Marchenko2012} may gap out the edge states near the graphene/contact interface, contributing a QSH-specific contact resistance which lowers the plateau conductance.

Nonlinear transport measurements provide some additional insight into the nature of backscattering in the edge states.  In both the QSH and CAF regimes, the nonlinear transport data is
invariant under simultaneous inversion of the carrier density and source-drain bias, $V_{SD}$ (Figs. 4e-f).  The data thus respects charge conjugation symmetry within the graphene, possibly implying that the inelastic processes probed by large $V_{SD}$ are native to the electronic system.  Notably, the nonlinear conductance is not invariant under reversal of source drain bias alone.  We can understand this lack of symmetry as a natural consequence of dissipative edge transport in our time-reversal noninvariant system, where, in contrast to topological insulators, the counterpropagating edge states can be spatially separated.  Within this picture, reversing $V_{SD}$ changes the current carried by the inner and outer counterpropagating edge states.  If dissipation differs between the two states on a single edge and the two physical graphene edges are inequivalent, reversing $V_{SD}$ can be expected to result in different conductance.

The current experiments present the first proof of the CAF-QSH crossover in monolayer graphene.  In addition, they enable the study of QSH physics in a versatile material platform, enabling new experiments.  Most importantly, the high-field graphene QSH system differs from the conventional TRS QSH state through the crucial role of interactions, which lead to the spontaneous breaking of spin symmetry that generates the gapped CAF edge states.  We note that in this paper we have discussed experimental results in the context of mean-field treatments of interactions in the graphene zero Landau level\cite{Kharitonov2011,Kharitonov2012}.  Crucially, this treatment neglects the potential for the spin-ferromagnetic (or CAF) order parameter to reconstruct near the sample boundary\cite{Fertig2006,Jung2009}, possibly leading to a qualitative change in the nature of the edge charge carriers.  These results should inspire more careful future work, both experimental and theoretical, to both understand the true nature of the edge states and to use them as a building block for realizing novel quantum circuits.

\subsection*{Methods}

\subsubsection{Sample Fabrication.} Samples consist of graphene-hBN-graphite stacks fabricated by a dry transfer process\cite{Hunt2013}.  Samples are annealed in H$_2$-Ar atmosphere at 350$^o$C\cite{Dean2010} after each transfer step and after patterning of contacts by standard electron beam lithography techniques.  Before measurement, residual debris from the fabrication process is swept off the graphene flake with an atomic force microscope tip operated in contact mode\cite{Jalilian2011,Goossens2012} (the evidence of this is visible in the AFM micrograph inset to Fig. 2b).  The methods used produce a random alignment angle between graphene and hBN flakes.  However, in the samples reported on in this paper, we do not see any evidence for moir\'{e}-induced band structure reconstructions\cite{Hunt2013}, suggesting that either the twist angle between graphene and hBN lattices is large, or the coupling weak.

\subsubsection{Conductance measurements} Conductance measurements were made using $\sim$300 Hz voltage bias, with $V_{ac}$=100$\mu$V.  The sample was immersed in $^3$He liquid at 300mK for all measurements excepts those shown in Fig. 4a, where the temperature is indicated, and those shown in Fig. 3, which were taken at 150mK with the sample immersed in $^3$He/$^4$He mixture.  The angle between the magnetic field and graphene plane was controlled by a mechanical rotator.  The sample was aligned using high density Shubnikov de Haas oscillations, ensuring reproducible alignment to better than .025$^o$ in the large tilt angle regime, $B_\perp\ll B_T$. For multiterminal devices (A and C), all measurements are done between two pairs of contacts (top configuration of Fig. 3a) unless otherwise indicated, ensuring that only two uninterrupted edges are being measured.

\subsubsection{Capacitance measurements}
To measure capacitance, we used a HEMT-based amplifier to construct a low temperature capacitance bridge-on-a-chip\cite{Ashoori1992}.  A schematic representation of the bridge geometry and electronics appears in Supplementary Figure S6.  In this geometry, an AC excitation on the graphene sample is balanced against a variable phase and amplitude excitation on a known reference capacitor, which is located near the sample.  The sample bias was 900$\mu$V for the data set in Fig. 1b and 100$\mu$V for that in Fig. 3a, both of which were measured at 78 kHz.  The signal at the input of the HEMT amplifier was first nulled by adjusting the reference excitation using a home-built dual-channel AC signal generator, after which data were acquired off-balance by monitoring the in- and out-of phase voltage at the balance point as a function of the applied DC sample bias.  Biasing the transistor amplifier raised the base temperature of the cryostat, so that the temperature was 400mK during acquisition of the data in Fig. 1b and 250 mK for Fig. 3a.

Extracting density of states information from capacitance measurements requires that the measurement frequency, $f$, be lower than the inverse charging time of the experimental system\cite{Goodall1985}.  Because incompressible regimes in quantum Hall samples are also highly resistive, contrast in the capacitance signal can be generated by the swing in sample resistance rather than density of states.  In this case, capacitance minima appear because the sample can no longer charge on timescales of $f^{-1}$.  Experimentally, such a transition is accompanied by a peak in the out-of-phase (dissipation) signal, which reaches a a maximum when the sample resistance $R\sim 1/fC$, where $C$ is the sample capacitance.   In the capacitance data shown in Figs. 1b and 3a, the dissipation signal is less than $\sim10\%$ of the capacitive signal, suggesting that the observed features are mostly due to density of states.  We note that whether the capacitance dip associated with the metallic, high field $\nu=0$ state is due to incompressibility or high bulk resistance is immaterial to the conclusions in the paper: in either case, any metallic transport should be via edge states.

\subsection*{Acknowledgements.} We acknowledge helpful discussions with D. Abanin, A. Akhmerov, C. Beenakker, L. Brey, L. Fu, M. Kharitonov, L. Levitov, P. Lee, and J. Sau.
B.H. and R.C.A. were funded by the BES Program of the Office of Science of the US DOE, Contract No. FG02-08ER46514 and the Gordon and Betty Moore Foundation through Grant GBMF2931.  J.D.S-Y, and P.J-H. have been primarily supported by the US DOE, BES Office, Division of Materials Sciences and Engineering under Award DE-SC0001819. Early fabrication feasibility studies were supported by NSF Career Award No. DMR-0845287 and the ONR GATE MURI. This work made use of the MRSEC Shared Experimental Facilities supported by NSF under award No. DMR-0819762 and of Harvard's CNS, supported by NSF under grant No. ECS-0335765. Some measurements were performed at the National High Magnetic Field Laboratory, which is supported by NSF Cooperative Agreement DMR-0654118, the State of Florida and DOE. AY acknowledges the support of the Pappalardo Fellowship in Physics.
\subsection*{Contributions}
AFY and JDSY conceived the experiment.  JDSY and SHC fabricated the samples. AFY, JDSY, and BH performed the experiments, analyzed the data, and wrote the paper.   TT and KW grew the hBN crystals.  RCA and PJH advised on experiments, data analysis, and paper writing.


%

\clearpage
\section{Supplementary Figures}
\begin{description}
\section*{Index of supplementary figures}
\item[Table 1] \hfill \\
Physical parameters of measured samples
\item[Figure S1] \hfill \\
Images of measured devices
\item[Figure S2] \hfill \\
Conductance as a function of $B_\perp$, $B_\textrm{T}$, and gate voltage for samples A, B and C.
\item[Figure S3] \hfill \\
$G_\textrm{cnp}$ as a function of $B_\perp$ and $B_{T}$  for samples A, B and E.
\newpage
\item[Figure S4] \hfill \\
Nonlocal measurements for sample C in the QSH and CAF regime.
\item[Figure S5] \hfill \\
Double conductance peaks in seven different samples.
\item[Figure S6] \hfill \\
Temperature dependence of the charge-neutrality point conductivity for Sample B.
\item[Figure S7] \hfill \\
Schematic of the capacitance bridge-on-a-chip in tilted magnetic field.
\item[Figure S8] \hfill \\
Tilted-field magnetotransport in zero-field insulating monolayer graphene.
\end{description}



\begin{center}

\begin{table*}[H!]
\begin{center}
	\caption{\label{tab:example}  \textbf{Physical parameters of measured samples}.  The studied devices consist of sequentially stacked flakes of thin graphite, h-BN, and monolayer graphene on an insulating Si wafer with 285nm of thermally-grown SiO$_{2}$.  The bottom graphite layer serves as a local gate electrode as well as to screen charge inhomogeneity in the graphene.  The table lists the details of the samples discussed in the main text (Samples A, B, and C), as well as for additional samples which are presented in these online supporting materials (Samples D-I).}
	\begin{ruledtabular}
	\begin{tabular}{c c  c c}
	\specialcell{ Sample \\ Name}   &\specialcell{ BN \\ Thickness (nm)}  &\specialcell{Sample \\ Dimensions LxW ($\mu$m)}    & \specialcell{Aspect Ratio \\ (L/W)}  \\
	 A & 14 &  3.6 x 1.4 & 2.6 \\  
	 B & 9 &  1.8 x 0.9 &  2\\ 
	 C & 8.4 &  1.2 x 1.2 & 1 \\ 
	 D & 4.2 &  2.1 x 0.7 & 3 \\ 
	 E& 3.4 &  1.6 x 0.7 & 2.2 \\ 
	 F & 4 & 1.1 x 0.9 &  1.2\\ 
	 G & 4.4 &   1 x 1 & 1\\ 
	 H & 2.2 &  1.4 x 2 &  0.7\\ 
	 I & 3.8 &  0.7 x 1.8 & 0.4\\ 
	
	\end{tabular}
	\end{ruledtabular}
	\label{sampletable}
\end{center}
\end{table*}
\end{center}

	\begin{figure*}[H!]
		 \renewcommand{\thefigure}{S\arabic{figure}}
	\begin{center}
	\includegraphics[scale=1.0]{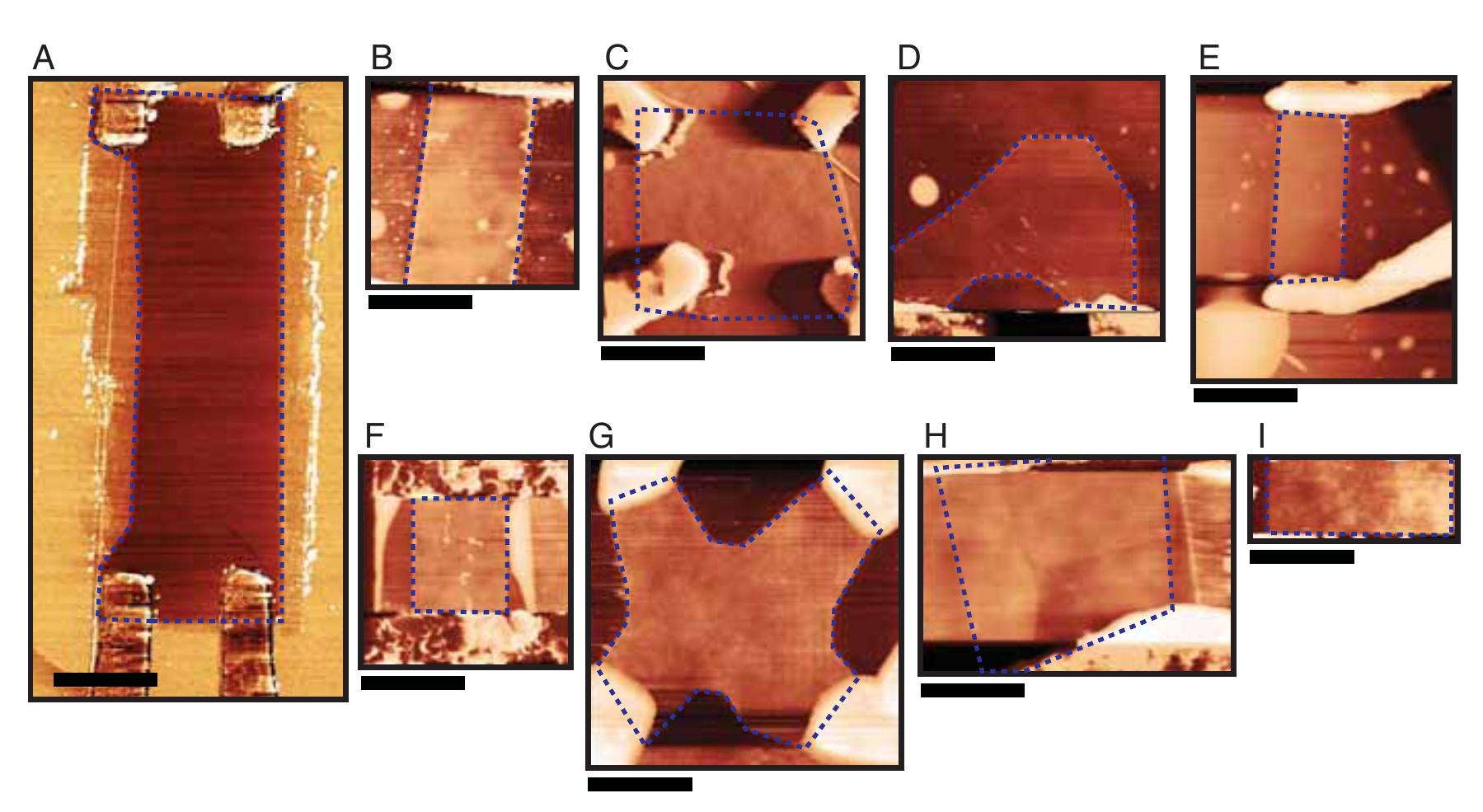}
		\caption{\textbf{Images of measured devices.}  False color AFM images of the devices enumerated in Table I.  Dashed lines outline the graphene boundary.  Black scale bars correspond to 1 $\mu$m.}
		\label{DeviceImages}
	\end{center}
\vspace{0mm}
\end{figure*}	


	\begin{figure*}[H!]
		 \renewcommand{\thefigure}{S\arabic{figure}}
	\begin{center}
	\includegraphics[scale=1.0]{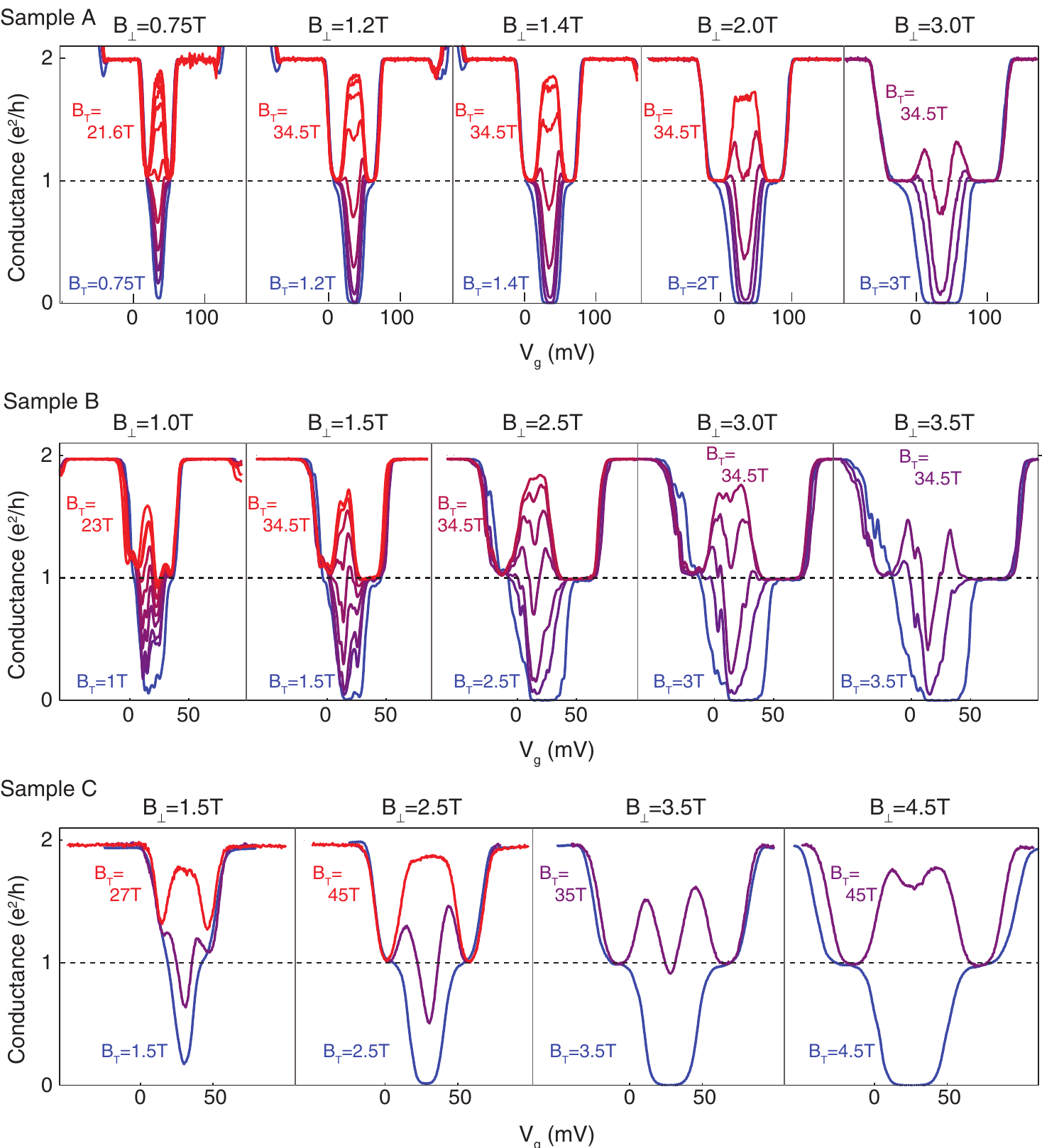}
		\caption{\textbf{Conductance as a function of $B_\perp$, $B_\textrm{T}$, and gate voltage for samples A, B and C.}   Coloring of lines from blue to red indicates increasing $B_\textrm{T}$, with $B_\perp$ as indicated at the top of each panel.}
		\label{moreTilts}
	\end{center}
\vspace{0mm}
\end{figure*}	

	\begin{figure*}[H!]
		 \renewcommand{\thefigure}{S\arabic{figure}}
	\begin{center}
	\includegraphics[scale=1.0]{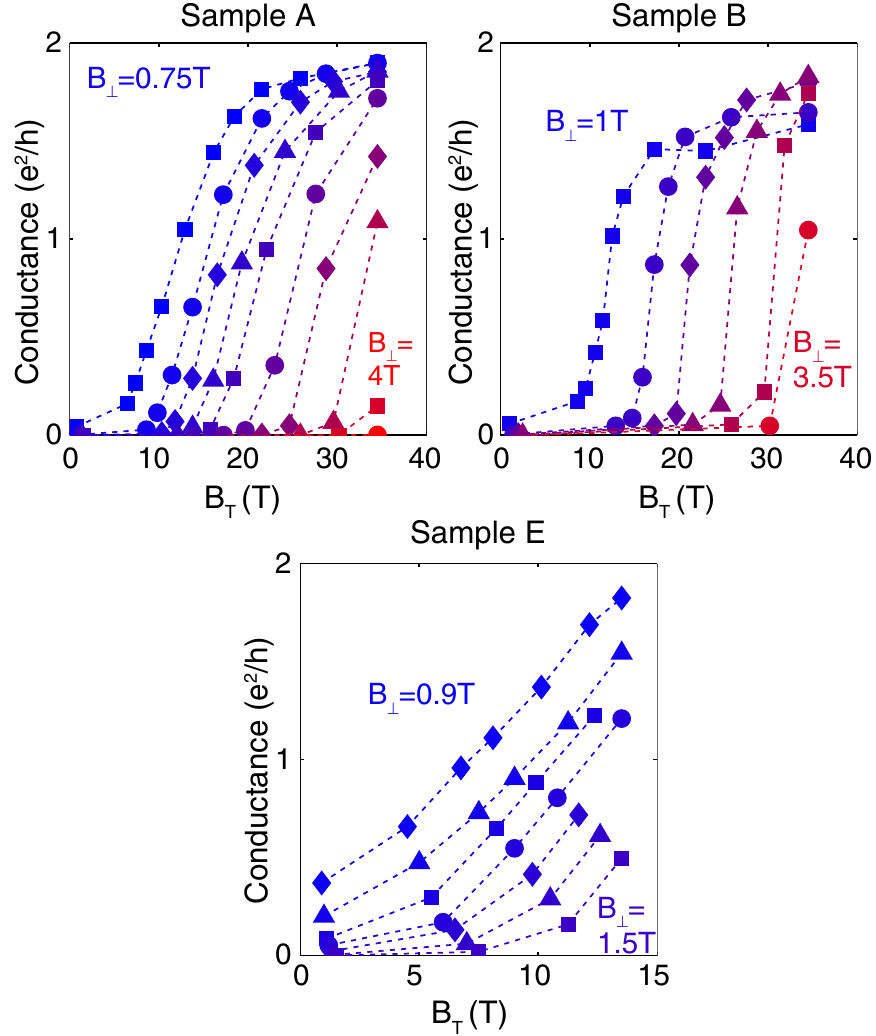}
		\caption{\textbf{$G_\textrm{cnp}$ as a function of $B_\perp$ and $B_{T}$  for samples A, B and E.}  Correspondingly higher values of $B_{T}$ are required to induce the transition for higher values of $B_\perp$.  For Sample A, the curves correspond to (blue to red) $B_\perp$=0.75, 1, 1.2, 1.4, 1.6, 2, 2.5, 3, 3.5, and 4T.  For Sample B, the curves correspond to (blue to red) $B_\perp$=1, 1.5, 2, 2.5, 3, and 3.5 T.  For Sample E, the curves correspond to (blue to red) $B_\perp$=0.9, 1, 1.1, 1.2, 1.3, 1.4, and 1.5T.}
		\label{Gcnp}
	\end{center}
\vspace{0mm}
\end{figure*}	


	\begin{figure*}[H!]
		 \renewcommand{\thefigure}{S\arabic{figure}}
	\begin{center}
	\includegraphics[scale=1.0]{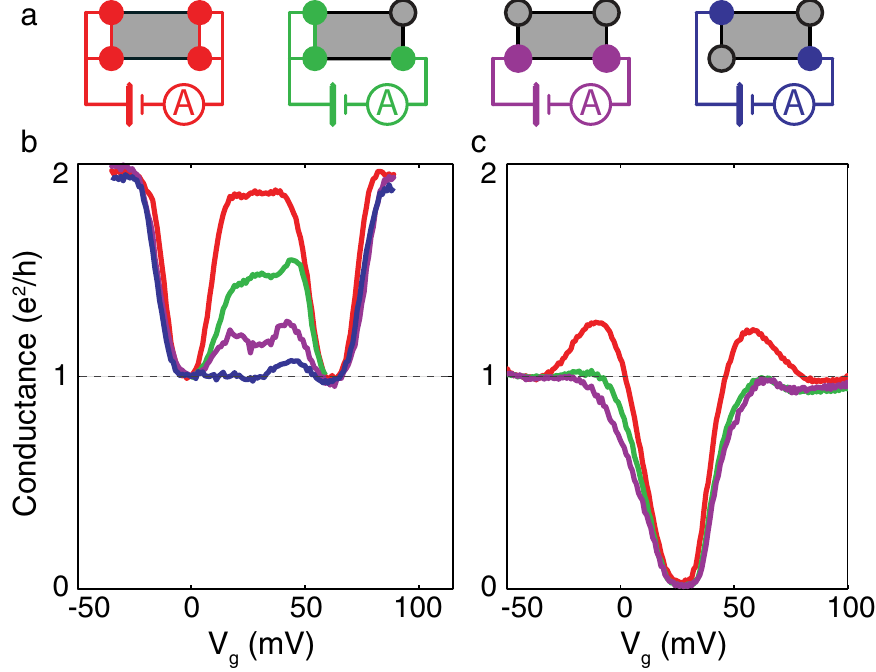}
		\caption{\textbf{Nonlocal measurements for sample C in the QSH and CAF regime.} In the main text, evidence for conduction via edge states in the QSH and CAF regimes is provided by nonlocal transport measurements in Sample A (Figures 2b and 4b).  Due to conduction through counter-propagating edge states, interrupting an edge with a floating contact decreases the 2-terminal conductance much more than would be expected in a  diffusive transport model.  Here we provide an additional example of this behavior for sample C.  \textbf{a,} Schematic of distinct 2-terminal measurement topologies with different number of floating contacts (hollow circles).  \textbf{b,} QSH regime, $B_{\perp}$=2.7T and $B_\textrm{T}$=45T.  \textbf{c,}  CAF regime, $B_{\perp}$=5.9T and $B_\textrm{T}$=45T.  Curves are color coded according to the measurement schematics, as in the main text. Due to a small gate leak in one of the contacts, these specific nonlocal measurements underestimate the conductance by a scale factor which was adjusted for by fitting the $\nu=-1$ plateau to a conductance of $e^2/h$.}
		\label{moreNonlocal}
	\end{center}
\vspace{0mm}
\end{figure*}	


	\begin{figure*}[H!]
		 \renewcommand{\thefigure}{S\arabic{figure}}
	\begin{center}
	\includegraphics[scale=1.0]{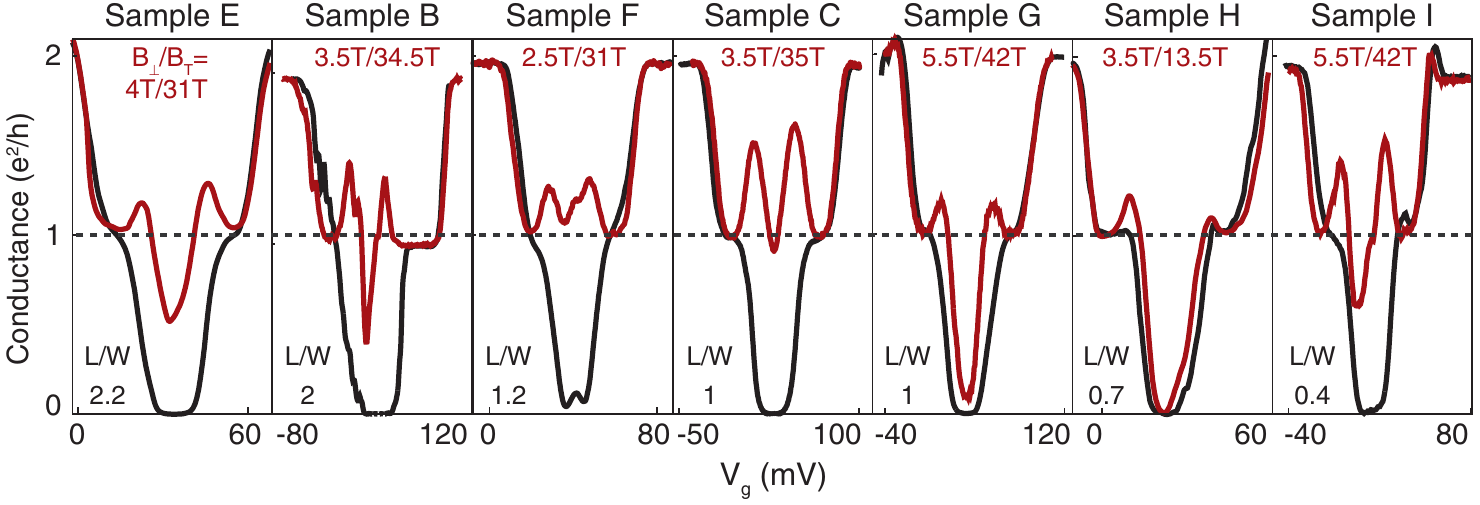}
		\caption{\textbf{Double conductance peaks in seven different samples.} A generic feature of the intermediate regime between the insulating and metallic QSH regimes is the appearance of double conductance peaks close to $\nu=0$. The figure shows two-terminal conductance vs. backgate voltage $V_G$.  Purely perpendicular magnetic field only ($B_\textrm{T}=B_\perp$, black lines) results in an insulating state at $\nu=0$.  Increasing the total magnetic field while keeping the perpendicular component constant ($B_\textrm{T}>B_\perp$, red lines), induces a transition to the CAF with associated double conductance peak feature.  Samples are ordered from left to right by descending aspect ratio.}
		\label{allLobes}
	\end{center}
\vspace{0mm}
\end{figure*}	



	\begin{figure*}[h]
		 \renewcommand{\thefigure}{S\arabic{figure}}
	\begin{center}
	\includegraphics[scale=1]{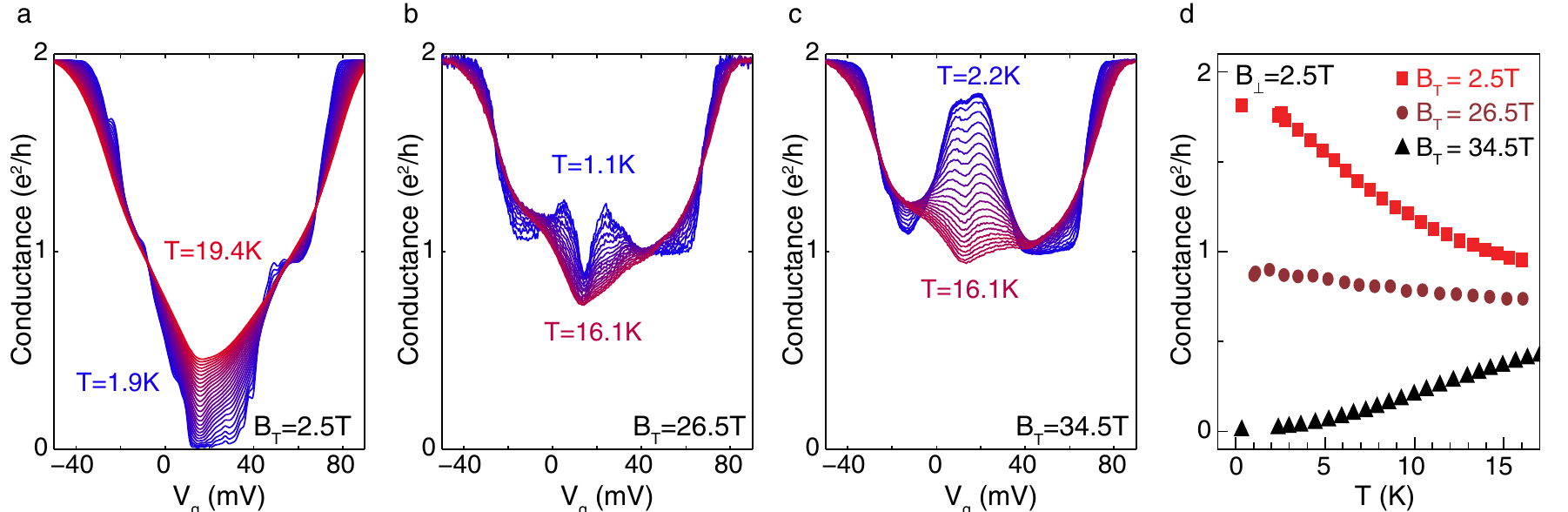}
		\caption{\textbf{Temperature dependence of the charge-neutrality point conductivity for Sample B.} \textbf{a-c,} Gate sweeps for sample B at constant $B_\perp$~=~2.5T and $B_\textrm{T}$ = 2.5T, 26.5T and 34.5T for a, b, and c, respectively.  \textbf{d,} Conductance at the charge neutrality point as a function of temperature for the data in a, b, and c.  A clear insulating dependence ($\partial G/\partial T>0$) is observed for $B_\perp=B_\textrm{T}$.  With increased $B_\textrm{T}$, in the intermediate regime, the double conductance peaks between $\nu=0$ and $\nu=\pm1$ display a weakly metallic temperature dependence ($\partial G/\partial T<0$) while $G_{cnp}$ is very weakly insulating.  In the QSH regime ($B_\textrm{T}$ >> $B_\perp$), where the conduction is along edge channels, the temperature dependence at $\nu=0$ is metallic. }
		\label{tempdepdipper}
	\end{center}
\vspace{-5mm}
\end{figure*}	


	\begin{figure*}[h]
		 \renewcommand{\thefigure}{S\arabic{figure}}
	\begin{center}
	\includegraphics[width=3.5 in]{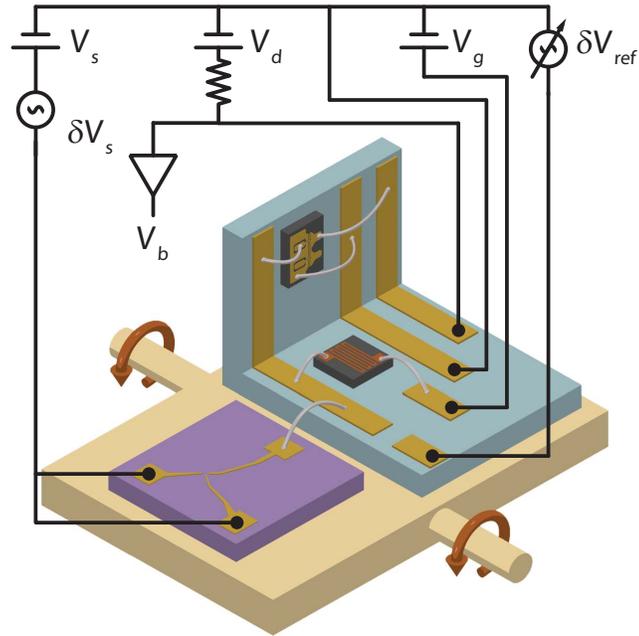}
		\caption{\textbf{Schematic of the capacitance bridge-on-a-chip in tilted magnetic field.} The magnetic field points up in the schematic.  Beige: sample stage, showing axis of rotation (red arrows).  Purple: graphene sample mount.  Blue: transistor mount with 90$^{\circ}$ bend.  The HEMT is mounted on the face angled 90$^{\circ}$ from the graphene sample mount and with the plane of its 2D conduction channel perpendicular to the sample stage axis of rotation.  A single wire bond connects the two mounts, from the graphite back gate to the balance point of the capacitance bridge.  The transistor is gated by applying $V_g$ to the balance point/graphite back gate through a 100 M$\Omega$ chip resistor.  Combined with total capacitance of the balance point to ground ($\sim3$ pF), this sets the low frequency cutoff for the measurement at $\sim1$ kHz.  The density of electrons in the graphene sample is determined by the DC voltage difference between the graphene sample and the graphite back gate, namely by $V_s-V_g$.  In the main text, this is compensated for, and all capacitance measurements are shown as a function of the graphite gate voltage relative to grounded graphene.  All components shown in black are at room temperature.}
		\label{capschematic}
	\end{center}
\vspace{-5mm}
\end{figure*}	
	\begin{figure*}[h]
		 \renewcommand{\thefigure}{S\arabic{figure}}
	\begin{center}
	\includegraphics[scale=1.0]{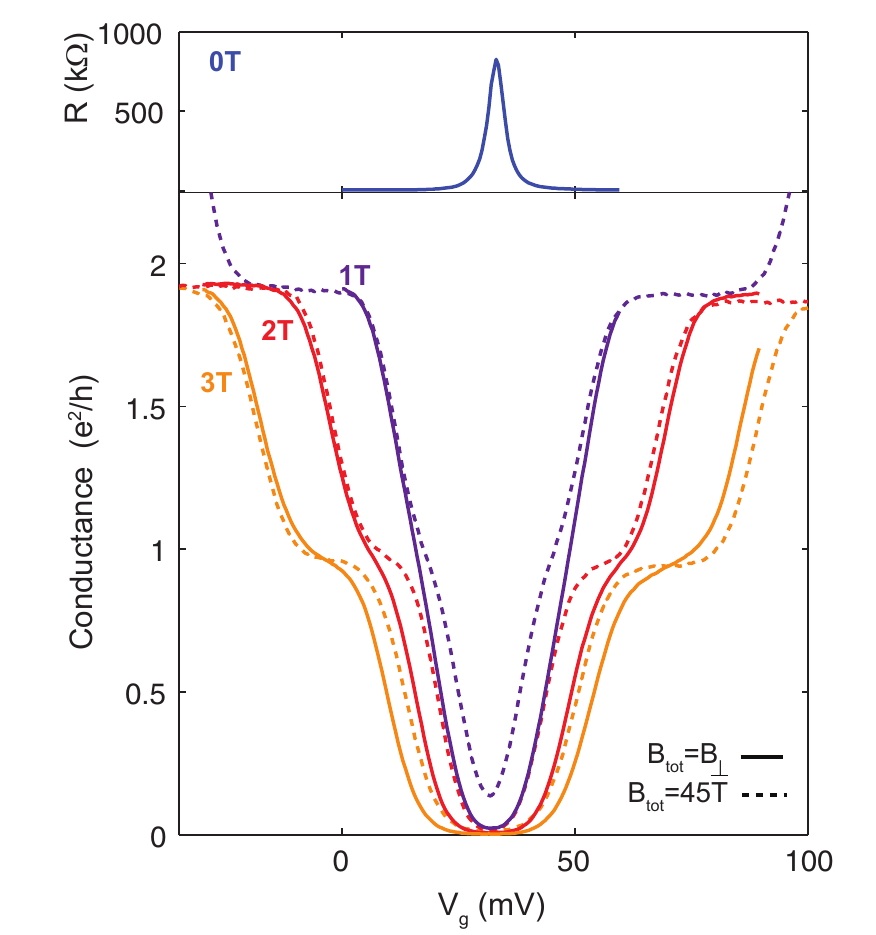}
		\caption{\textbf{Tilted-field magnetotransport in zero-field insulating monolayer graphene.}  In a fraction of devices having the identical geometry to those presented in the main text, we find that rather than a conductivity of $\sim e^2/h$ at charge neutrality these devices instead exhibit insulating behavior at the CNP at zero applied magnetic field.  We ascribe this insulating behavior to the opening of a bandgap at the CNP due to the effect of an aligned hBN substrate\cite{Hunt2013}.   The top panel shows resistance of the device in zero magnetic field. This device has a resistance of 825 k$\Omega$ at the CNP in zero magnetic field and T=.3K.  As with the devices described in the main text, the insulating state becomes stronger in a perpendicular magnetic field.  In the bottom panel, solid lines are gate sweeps at constant $B_{\perp}$=1, 2 and 3T and $B_\textrm{T}=B_{\perp}$.  Dashed lines are for the corresponding sequence of  $B_{\perp}$=1, 2 and 3T but with $B_\textrm{T}$=45T for each.  Data taken at 0.3K.   Semiconducting graphene samples do not show any sign of QSH-type physics, at least up to 45T.  Even for $B_{\perp}$=1T and $B_\textrm{T}$=45T, the conductance at the CNP increases only slightly, from 0.02 $e^2/h$ with zero in-plane field to 0.14$e^2/h$ with $B_\textrm{T}$=45T.   This is understandable, as even neglecting interaction effects, closing a moir\'{e}-induced band gap of $\Delta=10$ meV requires a Zeeman field of nearly $\Delta/(g \mu_B)\approx 85$~T. We note that in these samples, the ground state at $B_\textrm{T}=B_\perp$ may not be an antiferromagnet.}
		\label{tiltedinsulator}
	\end{center}
\vspace{0mm}
\end{figure*}	

\end{document}